\documentclass[aps,prb,showpacs,twocolumn,letterpaper,amsmath,amssymb]{revtex4}
\pdfoutput=1
\usepackage{graphicx}
\usepackage{dcolumn}
\usepackage{bm}

\begin{document}

\title{Optimization of Mn Doping in Group-IV-based Dilute Magnetic Semiconductors by Electronic Co-dopants}

\author{Hua Chen}
\author{Wenguang Zhu}
    \affiliation{Department of Physics and Astronomy, University of Tennessee, Knoxville, TN 37996}
    \affiliation{Materials Science and Technology Division, Oak Ridge National Laboratory, Oak Ridge, TN 37831}

\author{Efthimios Kaxiras}
 \affiliation{Department of Physics and School of Engineering and Applied Sciences, Harvard
University, Cambridge, MA 02138}

\author{Zhenyu Zhang}
 \affiliation{Materials Science and Technology Division, Oak Ridge National Laboratory, Oak Ridge, TN 37831}
 \affiliation{Department of Physics and Astronomy, University of Tennessee, Knoxville, TN 37996}
\date{\today}

\begin{abstract}
The percentage of substitutional doping of magnetic atoms (Mn) in group-IV-based dilute magnetic semiconductors (DMS) can be increased by co-doping with another conventional electronic dopant (e-dopant), as demonstrated from first-principles calculations recently [Zhu {\it et al.}, Phys. Rev. Lett. {\bf 100}, 027205 (2008)]. Here, we report extensive theoretical investigations of the kinetic and thermodynamic characteristics of several
co-doped systems including bulk Si and Ge as hosts and various group-III and group-V e-dopants. The main findings are as follows: The \textit{n-p} pairing of \textit{n}-type e-dopants with \textit{p}-type substitutional Mn is energetically stable in bulk Ge and Si. Mn atoms move from interstitial sites to substitutional sites easier (with lower kinetic barriers)  in the presence of a neighboring \textit{n}-type e-dopant. Magnetic coupling between two Mn atoms in bulk Ge oscillates between positive (ferromagnetic) and negative (antiferromagnetic) values with increasing Mn-Mn distance,
but in Mn/As co-doped Ge the coupling parameter remains positive at all distances beyond nearest-neighbors and this qualitative difference does not change with the doping level. For Mn doped Si, all coupling values except for the nearest neighbor one are positive and do not change much upon co-doping. We find an unconventional magnetic anisotropy in the co-doped system, that is, the dependence of magnetic coupling on the relative positions of the magnetic ions and their neighboring e-dopants. We map the calculated magnetic coupling to a classical Heisenberg model and employ Monte Carlo simulations to estimate the Curie temperature (T$_c$). We find that in Mn doped Ge no ferromagnetic order exists for Mn concentrations ranging from 3.13\% to 6\%. Instead, a spin-glass phase transition occurs at $\sim$5~K at 5\% Mn doping. For Mn/As co-doped Ge, T$_c$ increases nearly linearly with the Mn concentration and reaches 264~K at 5\% Mn doping.

\end{abstract}


\maketitle

\section{\label{sec:intro}Introduction}

Dilute magnetic semiconductors (DMS) have attracted much interest in the condensed matter community not only because of their promising application in spintronic devices~\cite{wolfscience01,zuticrmp04}, but also because of the many new and important theoretical issues which arise from the study of this unique class of disordered magnetic system \cite{ohnoscience98,dietlscience00,bhattjs02,timmjpcm03,jungwirthrmp06}. As for specific materials, besides the most extensively studied (III,Mn)V systems \cite{jungwirthrmp06,sanvitojs02}, Mn doped group-IV semiconductors such as Ge and Si also show promise for real applications\cite{parkscience02,ohtapb01,nakayamapb01,abeass99,picozziprb04,bolducprb05,zengprl08}.
In order to realize this promise, a Curie
temperature comparable to room temperature or higher is required.
Both theory and experiment indicate that the Curie temperature of the above-mentioned materials is exceptionally sensitive to the ratio of interstitial to substitutional Mn atoms
\cite{wuprl05, erwinprl02, yuprb02, haoprl07, timmjpcm03}.
In (III,Mn)V as well as (Mn,IV) systems, substitutional Mn atoms act as acceptors and provide holes which,
according to current understanding, are the mediator of magnetic interactions between magnetic moments in these materials. Interstitial Mn atoms~\cite{yuprb02} are identified to be donors, and tend to compensate the holes and magnetic moments induced by the substitutional Mn \cite{jungwirthrmp06}. Furthermore, though annealing is an effective way to decrease the percentage of
interstitial Mn while keeping the homogeneity in (Ga,Mn)As,~\cite{kuapl03,edmondsprl04} it is less useful for Mn$_{x}$Ge$_{1-x}$ and Mn$_{x}$Si$_{1-x}$,~\cite{doraziojmmm03,zengprb04,jametnm06,kwonssc05} which makes it very difficult to get high quality samples of these materials using conventional methods.

In our recent work\cite{zhuprl08}, a novel way to enhance the
substitutional doping of Mn in Ge and Si was proposed. In this
method, an additional conventional electronic dopant (e-dopant) such as As or P is introduced in the doping process. Using first-principles electronic structure calculations, we
were able to show that the co-doping approach can substantially lower the
energy of Mn atoms at substitutional sites relative to that at interstitial sites, as well as the
energy barrier which the Mn atoms have to overcome in order to be incorporated into
substitutional sites. In addition, the assisting e-dopant
enhances the magnetic coupling between substitutional
Mn atoms. A new type of magnetic
anisotropy was also found, which depends on the proximity of the
assisting e-dopant to a Mn dopant, rather than the
direction of magnetic moments relative to the lattice direction of the host or to
other moments.

In this paper, we present a detailed \textit{ab initio} investigation of this novel approach by analyzing the kinetic and thermodynamic issues related to the stability of various dopant-host combinations.  We then calculate the magnetic coupling between two Mn atoms in bulk Ge and Si, and find that in Ge, the coupling oscillates between positive (ferromagnetic) and negative (antiferromagnetic) values with the Mn-Mn distance. But in Mn/As co-doped Ge the coupling parameter remains positive at all distances beyond nearest-neighbors, and this qualitative difference does not change with the doping level. For Mn doped Si, all the couplings except for the nearest neighbor one are positive and do not change much upon co-doping. We also carry out Monte Carlo simulations to obtain the Curie temperatures of the co-doped materials. We find that in Mn doped Ge no ferromagnetic order exists for Mn concentrations ranging from 3.13\% to 6\%. Instead, a spin-glass phase transition occurs at $\sim$5~K at 5\% Mn doping. For Mn/As co-doped Ge, T$_c$ increases nearly linearly with the Mn concentration and reaches 264~K at 5\% Mn doping.

The paper is organized as follows: In section \ref{sec:methods} we present the
computational methodology, including details of our
\textit{ab initio} treatment and the Monte Carlo simulations. The main
\textit{ab initio} results are given in Sec.\ref{sec:dft}, where the kinetic and energetic properties of various combinations of host materials (Ge, Si)
and assisting dopants (As, P, Al, Ga) are investigated. Magnetic
interactions in As co-doped Mn$_{x}$Ge$_{1-x}$ are discussed in Sec.\ref{sec:magnetism}.
In Sec.\ref{sec:tc} the
\textit{ab initio} results of magnetic coupling are used to find the
transition temperature T$_c$ of Mn/As co-doped Ge. The
discussion and summary are provided in the last two sections.

\section{\label{sec:methods} Methods}

Our spin-polarized first-principles calculations are carried out
using the Vienna \textit{ab initio} simulation package (VASP)
\cite{kresseprb96}, a density functional theory approach using the
projector augmented wave (PAW) method \cite{blochlprb94, kresseprb99} and the generalized
gradient approximation (PBE-GGA) \cite{perdewprl96} for exchange-correlation.
A default plane-wave energy cutoff of 269.9 eV is consistently used in
all Mn calculations.
These choices produce a bulk Ge and Si lattice constants of 5.78 \AA{}(experimental value\cite{hubbardjac75} 5.66 \AA{}) and 5.47 \AA{} respectively
(experimental value\cite{singhac68} 5.43 \AA{}).

In our calculations of the co-doping process the supercell
size is chosen to be a $2 \times 2 \times 2$ multiple of the conventional
cubic cell of the diamond lattice which contains 8 atoms. Hence, there are 64
atoms in one supercell, and one of them is replaced by an Mn atom,
corresponding to 1.56\% Mn concentration, comparable to what was
achieved experimentally \cite{parkscience02, liapl05, massc06}.
Different supercell sizes were used to study the
dependence of calculated results on Mn concentration.
Specifically, we used a $3 \times 3 \times 3$ supercell, which corresponds
to 216 atoms, and with one of them replaced by a Mn the concentration is 0.46\%.
In each calculation of the magnetic coupling between Mn atoms, two Mn
atoms are placed in a $3 \times 3 \times 3$ supercell, corresponding to
a 0.926\% Mn concentration.
We also selectively use a $2 \times 2 \times 2$ supercell for the magnetic
coupling with two Mn atoms in the supercell, corresponding to 3.125\% Mn,
for comparison. This setup is similar to previous
studies of Mn-Mn interactions in pure semiconductors\cite{zhaoprl03, stroppaprb03, wengprb05, mahadevanprl04}.

A uniform $4 \times 4 \times 4$ ($2 \times 2 \times 2$)
mesh, including the $\Gamma$ point (0, 0, 0), is chosen for Brillouin
Zone sampling in the $2 \times 2 \times 2$ ($3 \times 3 \times 3$) supercell. Optimized
atomic geometries are obtained when the forces on all the
unconstrained atoms are smaller in magnitude than 0.01 eV/\AA . The
\textquotedblleft{}climbing image Nudged Elastic Band" (NEB) method
\cite{henkelmanjcp00} is used to locate the transition state geometries
for the calculation of activation energy barriers. Typically four slab replicas between the initial and final geometries are enough to produce a smooth minimum energy path.

For the Monte Carlo simulations we use the Metropolis algorithm\cite{binder} and the
magnetic energy of the system is calculated using the classical
Heisenberg model, in which each magnetic ion is treated as a classical
moment and is placed at a randomly chosen site of the
supercell. The magnetic coupling parameters are extracted from
\textit{ab initio} results of the energy difference between parallel
and antiparallel spin configurations of two Mn moments at different
separations. At each temperature we use 50000 Monte Carlo steps per
moment for the system to relax, and calculate the thermal average in the
following 50000 steps. To determine the Curie temperature, we adopt the
fourth order cumulant crossing method based on finite-size scaling theory proposed by Binder
\cite{binderprl81,binder}. In applying this method we choose three supercell sizes:
$8 \times 8 \times 8$, $10 \times 10
\times 10$, $12 \times 12 \times 12$, and
40 configurations in each case for averaging.

\section{\label{sec:dft} \textit{Ab initio} study of the co-doping processes}

\subsection{\label{sec:dft-pure} Study on intrinsic (Mn,IV) without additional dopants}

We first consider the equilibrium structure of a
single Mn dopant atom in bulk Si and Ge, and address the difficulty of lowering the
percentage of interstitial Mn impurities. A complete
understanding of the microscopic doping process requires detailed
knowledge of the energetics as well as the kinetics of dopants
in the host material\cite{erwinprl02}. In fact, an in-depth understanding of
the growth kinetics is particularly important, because DMS systems
are typically in a metastable state, since they are grown by co-doping the magnetic
dopants and host semiconductor atoms
using molecular beam epitaxy under nonequilibrium conditions \cite{ohnoapl96,ohnojap99,parkscience02,liapl05}.

\begin{figure}
 \includegraphics[width=3.4in]{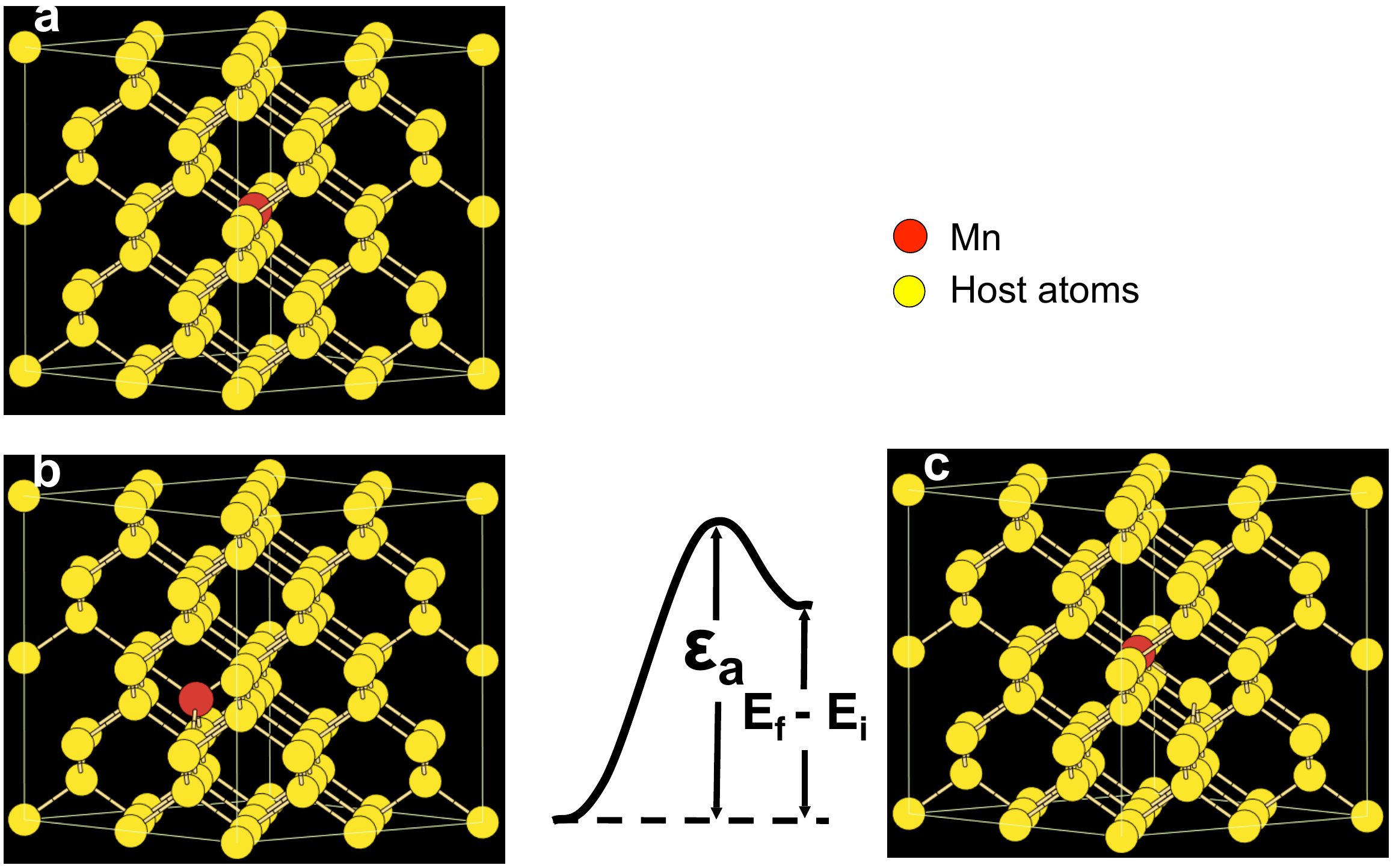}
 \caption{\label{fig:pure}\textbf{(color online)} Different Mn sites
 in bulk Si or Ge: (a) Mn at a substitutional site.
 (b) Mn at an interstitial site.
 (c) Final state of an interstitial Mn kicking out a neighboring host atom to an interstitial site and occupying the left-behind substitutional one. \label{fig:mnsites}}
\end{figure}

To address these issues,
we calculate the relative formation energy of a substitutional Mn(Fig.~\ref{fig:pure}(a)) and interstitial Mn(Fig.~\ref{fig:pure}(b)) atom in Ge and Si separately, which is defined as:
\begin{eqnarray}
\Delta E_1  = (E_{subst}  + \mu _{host} ) - E_{inter},
\end{eqnarray}
where $\mu _{host} $ is the host material's chemical potential. For Ge, our calculation gives $\Delta E_1  =  -0.63$eV.
Thus, in Ge the substitutional sites have a relatively
lower energy and are preferred by Mn atoms. However, for Si the opposite is true and
$\Delta E_1 =  + 0.58$. This reversed site preference~\cite{dasilvaprb04} makes it extremely
hard to achieve experimentally even a nominal concentration of substitutional Mn in
silicon.

We next consider kinetic aspects of the Mn doping process.
In order to get a high ratio of
substitutional to interstitial Mn, the process that an interstitial
Mn kicks out a host atom and becomes substitutional must take place more
often than the reverse process. Accordingly, we calculate the energy
difference between the initial (interstitial, (\ref{fig:pure})(b)) and final (substitutional, (\ref{fig:pure})(c))
states of this process: $\Delta E = E_f  - E_i $, and the energy barriers
$\varepsilon _a$ and $\varepsilon '_a $ for the reverse process.
Our calculation shows that for both Ge and Si
$\Delta E$ is positive (0.82 eV and 2.03 eV
respectively). This energy cost for the transition from initial to
final state defines the lower bound of the activation energy barrier
for the exchange process, which must be lower than $\sim 0.8$ eV
for efficient incorporation ( with a standard attempt frequency
10$^{12}$ sec$^{-1}$). Moreover, the actual energy barrier
$\varepsilon _a $ in either case (1.12eV for Ge and
$>$2eV for Si) is higher than the barrier of the
reverse process, which is calculated as $\varepsilon '_a
$=$\varepsilon _a $-$\Delta E$, with the latter being lower than 0.8 eV, further
facilitating the reverse processes. Thus, kinetically Mn is more stable
at interstitial sites rather than at substitutional sites in both Ge
or Si. In the following section we will address the issue of doping Mn together with another \textit{n-}
(P, As) or \textit{p}-type (Al, Ga) conventional e-dopant in order to explore
how the assisting dopants influence this site preference both
energetically and kinetically.

\subsection{Energetic and kinetic study on the co-doped systems}

Substitutional Mn in Ge is a \textit{p}-type double acceptor~\cite{picozzipssa06}.
Our proposal for a co-doping mechanism is based on the fact that the
electrostatic interaction between a \textit{n}-type and a \textit{p}-type dopant in a semiconductor
is attractive because of their different charge states (see below).
Thus an \textit{n}-type e-dopant may help to stabilize substitutional Mn atoms.

We start by noting that in Ge or Si there are two kinds of
interstitial sites: the hexagonal interstitial site $I_H$,
which has six nearest neighbors, and the tetrahedral site
$I_T $ with four nearest neighbors. Using
first-principles calculations we find that in \textit{n}-type doped Ge and Si,
the energy of a Mn sitting at the $I_H $ site is
different from that at the $I_T $ site. For P, As and Sb
doped Ge, the energy differences are 0.14, 0.09 and 0.04eV,
respectively, where a positive sign means the $I_H$
occupation has lower energy and is preferred. In
the case of either \textit{n}- or \textit{p}-doped Si as well as \textit{p}-doped Ge,
$I_T $ is preferred to $I_H $.
This dopant dependent preference can be qualitatively explained by the
local strain effect. Namely, a Mn atom and an \textit{n}-type dopant favor a relatively short
bonding distance, which is accommodated by
Mn occupying the $I_H $ site rather than the
$I_T $ site in Ge (the $I_H $
site has a shorter distance to its nearest neighbors than the
$I_T $ site). To show that this is indeed the case, we
reduce the lattice constant of Ge to the value of Si and calculate the
energy difference again. Then the results show that the preference
for Mn is changed to the $I_T $ site, because in this case
the distance between the $I_H $ Mn and \textit{n}-type
dopant becomes too short (compressive), whereas at the $I_T$
site the Mn/\textit{n}-type dopant bond length is close to its optimal value. We have
also checked to confirm that if we increase the lattice constant of Si
to that of Ge, the preference for Mn is changed to the $I_H$
site for the \textit{n}-type doped systems.

\begin{figure}
 \includegraphics[width=3.2in]{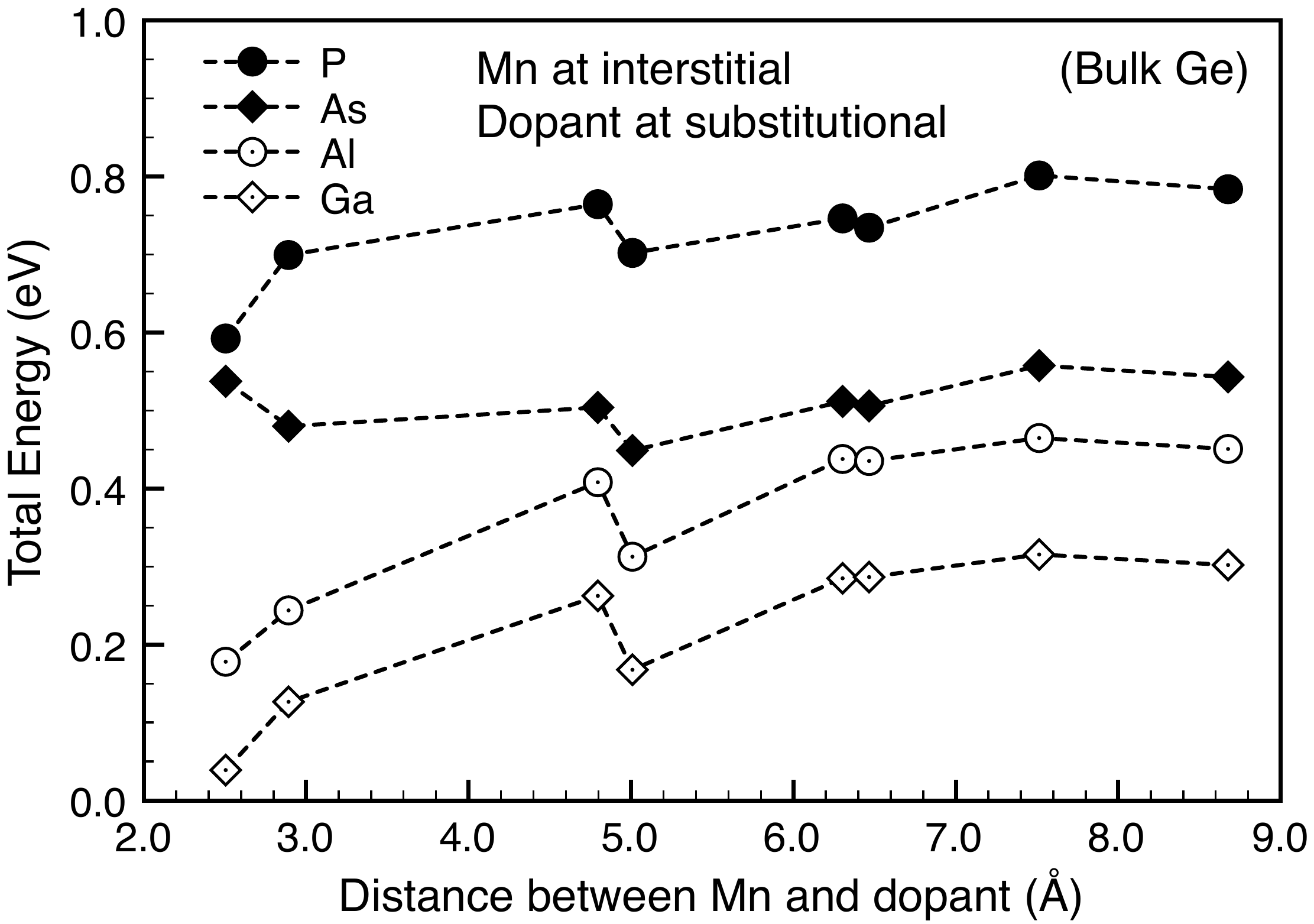}
 \caption{Calculated relative total energy as a function of the distance between an interstitial Mn and a substitutional dopant in bulk Ge.\label{fig:i-s-distance}}
\end{figure}

In the following we examine two possible kinetic processes of an
interstitial Mn atom becoming substitutional.
These processes share the same initial state with Mn occupying either
the $I_H $ or $I_T $ sites with a neighboring \textit{n}-type or \textit{p}-type e-dopant.
From our calculation of the total energy of a Ge-supercell with an interstitial Mn
and a substitutional e-dopant as a function of their separation,
shown in Fig.~\ref{fig:i-s-distance}, we find that shorter separation
is energetically preferred. Thus,
the choice of neighboring Mn/e-dopant pair configuration is reasonable.

\begin{figure}[b]
 \includegraphics[width=3.4in]{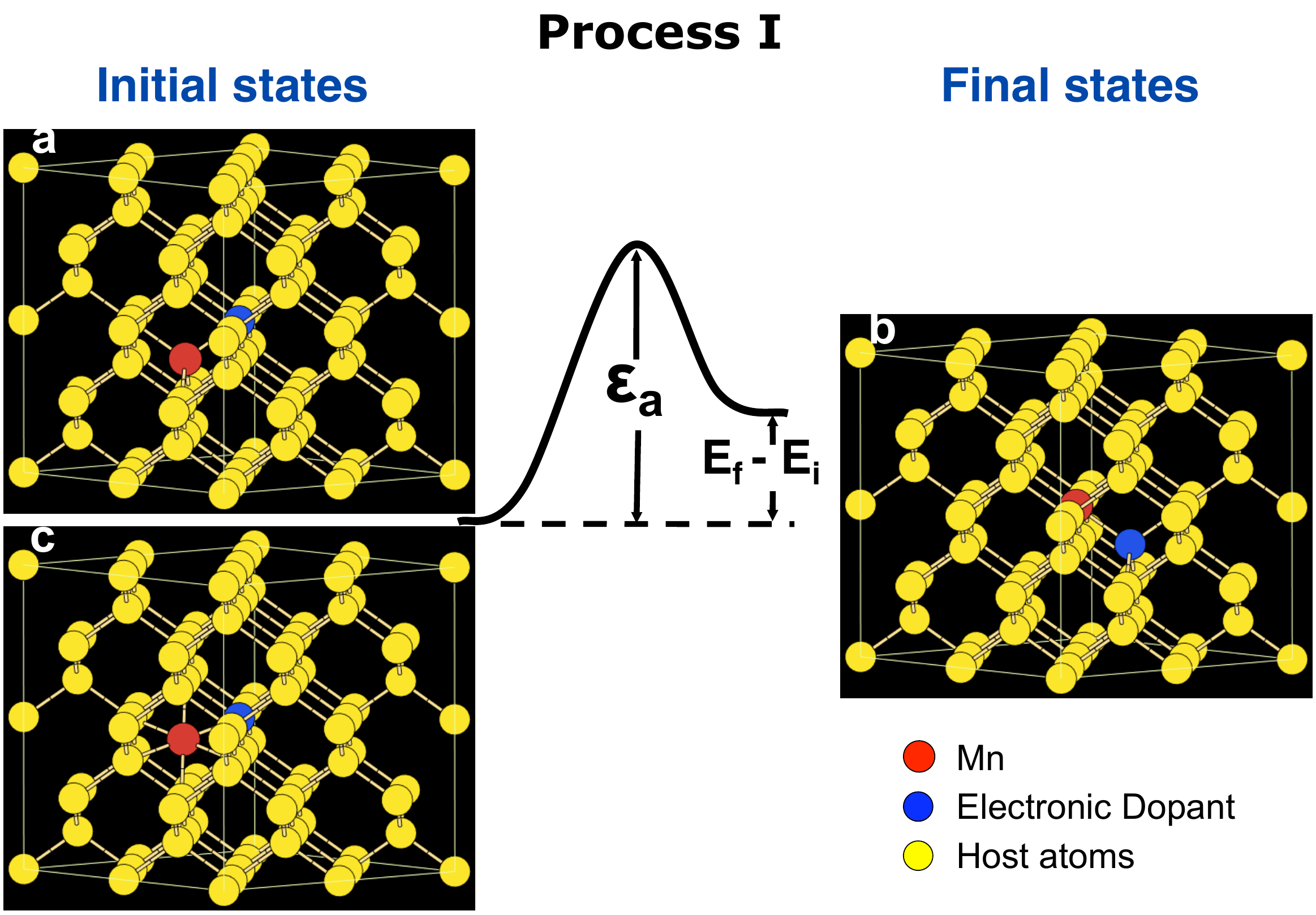}
 \caption{\label{fig:proc1}\textbf{(color online)} Atomic structures and schematic energy profiles
 for Process I:
 (a) Initial state with Mn in the $I_{T}$ position (except for $n$-type doped Ge);
 (b) Final state;
 (c) Initial state for $n$-type doped Ge with Mn at the $I_{H}$ position.}
\end{figure}

In the first process, denoted as Process I, we consider an interstitial Mn directly
exchanging position with its substitutional e-dopant neighbor.
In the final state, the e-dopant is pushed to an
adjacent interstitial site and the Mn atom moves into the
substitutional site left behind, as shown in Fig.~\ref{fig:proc1}.
Table \ref{tab:kinetics} summarizes the calculated
energy differences $\Delta E$ between the final and
initial states for \textit{n}-type and \textit{p}-type dopants in Si and Ge.
We find that only the P or As doped Ge (with
$\Delta E$=0.33 eV and 0.42 eV respectively) can
fulfill the requirement that $\Delta E<$0.8 eV. However, further examination of the
activation energy barriers for incorporation in these two cases gives
$\varepsilon _a $= 0.88 eV and 0.98 eV
respectively, which means this process is unlikely to happen in both
cases. Moreover, the reverse processes with $\varepsilon '_a$= 0.55 eV
and 0.56 eV, respectively, are more likely to occur.

\begingroup
\begin{table*}
\caption{Calculated energy differences $\Delta E=E_f-E_i$ (in eV) between
the final and initial states of Process I and Process II, illustrated in Fig.1.
$\varepsilon_a$ (in eV) is the activation
energy for a transition from the initial to final state. Results highlighted in bold correspond to processes for which $\Delta E$ or $\varepsilon_a$ or both are $< 0.8$ eV. All the results are for Mn
concentration of 1.56\%; results for selected cases with Mn concentration of 0.46\% are given in brackets.\label{tab:kinetics}}
\begin{ruledtabular}
\begin{tabular}{cccccccc}
\multicolumn{3}{c}{\textbf{Bulk Si}}&\multicolumn{5}{c}{\textbf{Bulk Ge}}\\
\cline{1-3} \cline{4-8}
\textbf{X}&\multicolumn{2}{c}{\textbf{$\Delta E=E_f-E_i$}}&\textbf{X}&\multicolumn{2}{c}{\textbf{$\Delta E=E_f-E_i$}}&\multicolumn{2}{c}{\textbf{$\varepsilon_a$}}\\
&\textbf{Proc.I}&\textbf{Proc.II}&&\textbf{Proc.I}&\textbf{Proc.II}&\textbf{Proc.I}&\textbf{Proc.II}\\
\textbf{Si}&2.03&&\textbf{Ge}&0.82[1.46]\\
\textbf{P}&1.46&0.89&\textbf{P}&\textbf{0.33}[\textbf{0.59}]&\textbf{0.03}[\textbf{0.17}]&0.88&\textbf{0.34}\\
\textbf{As}&1.55&1.09&\textbf{As}&\textbf{0.42}[\textbf{0.66}]&\textbf{0.05}[\textbf{0.34}]&0.98&\textbf{0.25}\\
\textbf{Al}&1.24&2.05&\textbf{Al}&0.94&1.54\\
\textbf{Ga}&1.79&2.43&\textbf{Ga}&1.05&1.52\\
\end{tabular}
\end{ruledtabular}
\end{table*}
\endgroup

\begin{figure}
 \includegraphics[width=3.2in]{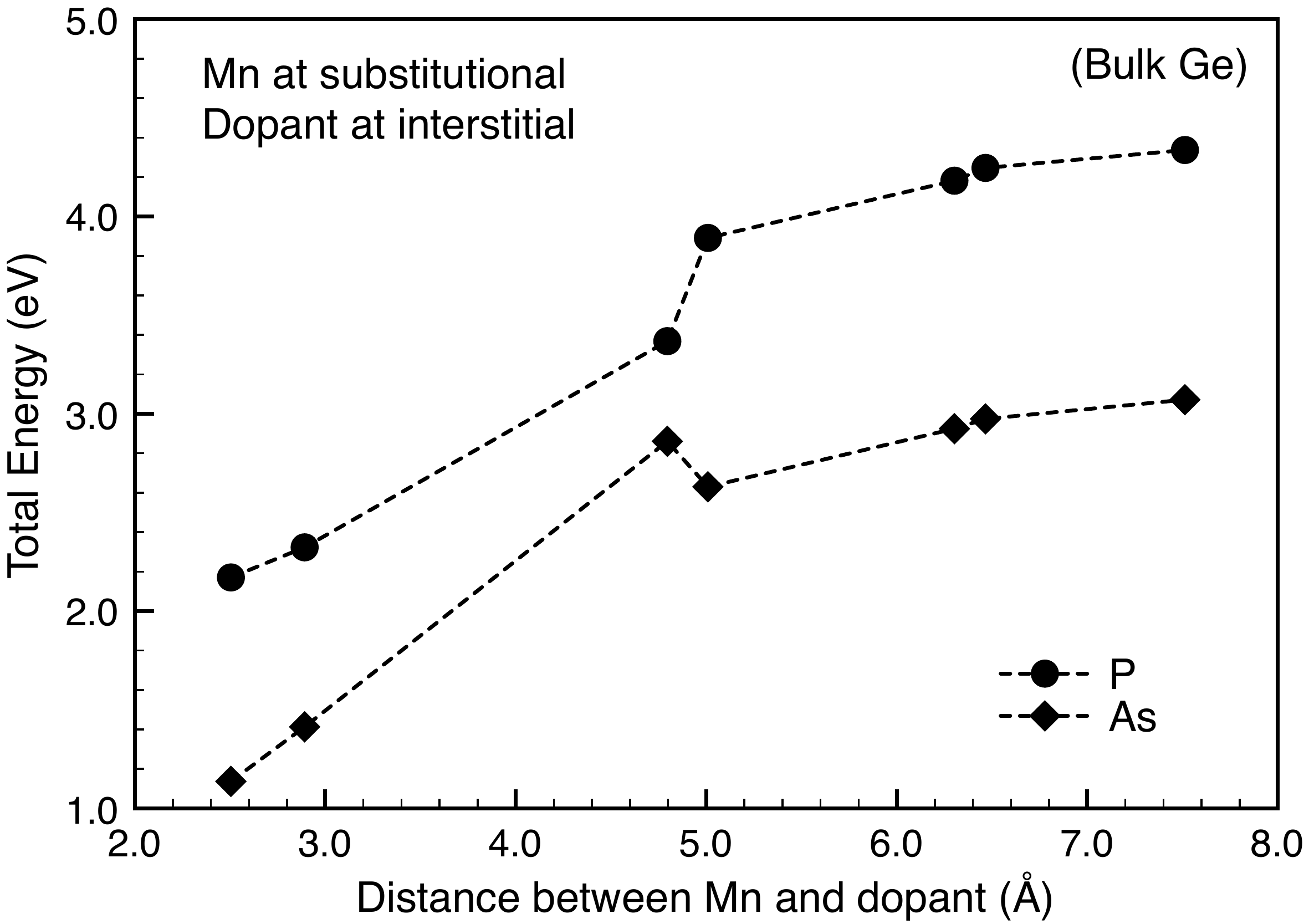}
 \caption{Calculated relative total energy as a function of the distance between a substitutional Mn and an interstitial $n$-type dopant in bulk Ge.\label{fig:s-i-distance}}
\end{figure}

Nevertheless, there is one possibility for the Mn atom to stay at
the substitutional site, that is, the kicked-out dopant atom diffuses
away rapidly so that the reverse process cannot happen.
This is ruled out by our calculation of the energy of a Mn/e-dopant
pair as a function of their separation, shown in Fig.\ref{fig:s-i-distance},
which shows that the e-dopant cannot diffuse away because the energy
increases with increasing separation.

\begin{figure}
 \includegraphics[width=3.4in]{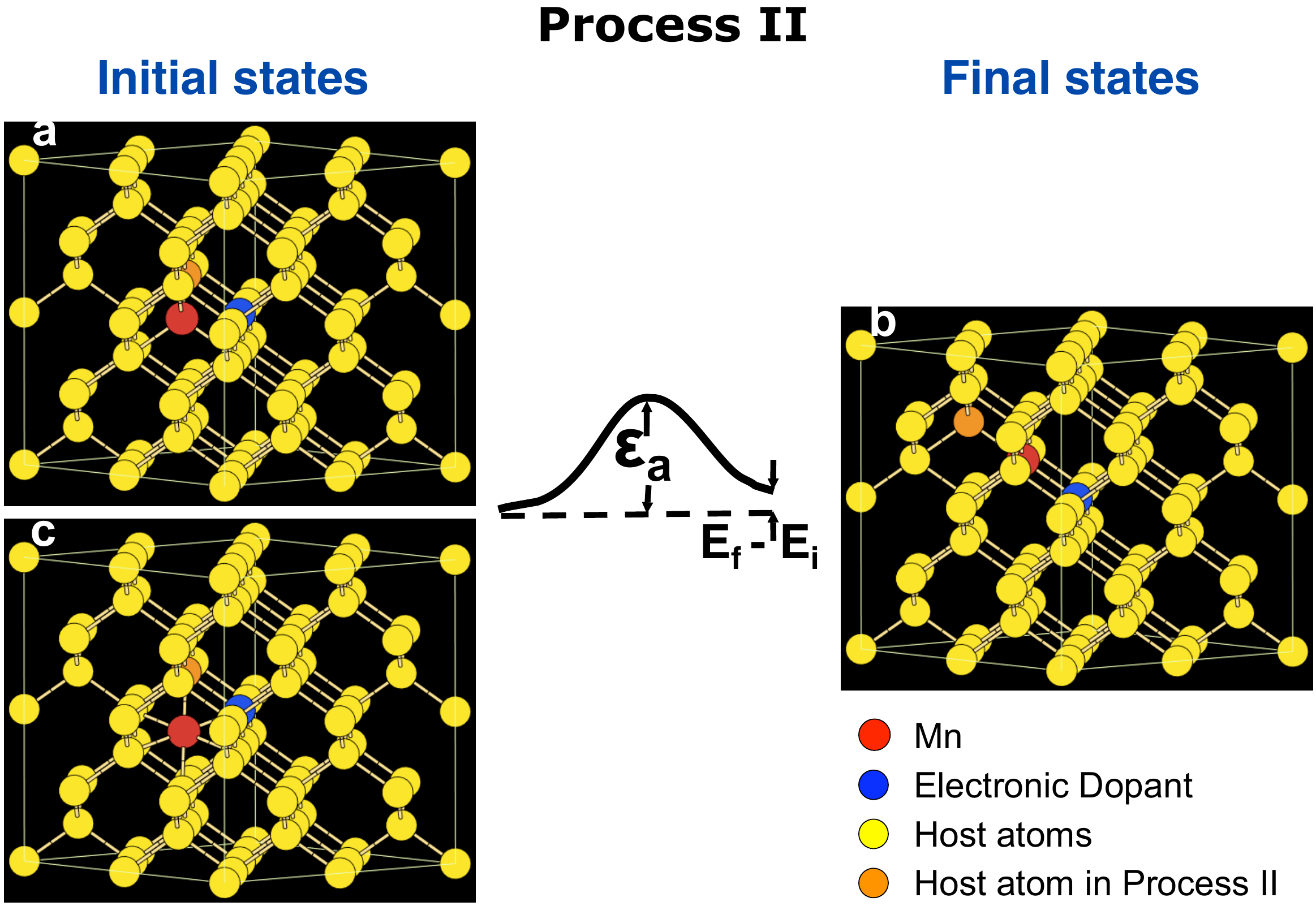}
 \caption{\label{fig:figure1}\textbf{(color online)} Atomic structures and schematic energy profiles of
 Process II:
 (a) Initial state with Mn in the $I_{T}$ position (except for $n$-type doped Ge);
 (b) Final state; (c) Initial state for $n$-type doped Ge, with Mn at the $I_{T}$ position.\label{fig:proc2}}
\end{figure}

We then consider a different process (Process II), which starts
from the same initial configuration as in Process I, but instead of exchanging
with the e-dopant, the Mn atom now pushes out a host atom next to the e-dopant to an interstitial site, and then occupies the substitutional site left behind. The final state is
shown in Fig.~\ref{fig:proc2}, in which the kicked-out interstitial host atom, the substitutional Mn
and the e-dopant are nearly collinear. The calculated
$\Delta E$ and $\varepsilon _a$ for various \textit{n}-type and \textit{p}-type
dopants in Si and Ge is also
shown in Table \ref{tab:kinetics}.  For \textit{n}-type doped Ge these
values are substantially lower than in Process I, and considerably
below the threshold value of 0.8 eV. For P and As doped Ge,
$\Delta E$ is actually quite low. Furthermore, the
activation barriers $\varepsilon _a $ for all the
three \textit{n}-type dopants is lees than 0.4eV. Qualitatively, this
substantial change in the energetic and kinetic characters originates
from the electrostatic attraction between the Mn atom, which behaves like
a \textit{p}-type dopant, and \textit{n}-type
dopants. Therefore, Process II, leading to substitutional Mn
atoms proximate to \textit{n}-type dopants, is more likely to happen in reality.

One issue that arises at this stage is whether the final state is thermodynamically stable.
To address this question,
we calculate the energy difference between interstitial Mn and substitutional Mn defined as
\begin{eqnarray}
\Delta E_2  = (E_{pair}  + \mu _{host} ) - E_{inter}.
\end{eqnarray}
Here $E_{inter} $ is the total energy of a
Mn/e-dopant pair, with the Mn sitting at an interstitial site, while
$E_{pair} $ is that with the Mn occupying a
substitutional site. The calculated interstitial-substitutional energy difference $\Delta E_2$ is shown in Table \ref{tab:energetics}. Compared to the results without \textit{n}-tyep dopants
in Sec.~\ref{sec:dft-pure}, the substitutional Mn in Ge becomes much more stable with the neighboring \textit{n}-type dopant. Moreover, the site preference of Mn in Si is reversed from interstitial to substitutional.

\begin{table}[b]
\caption{Relative formation energy of substitutional and interstitial Mn in the
presence of a neighboring substitutional $n$-type dopant,
defined as: $\Delta E_2=(E_{pair}+\mu_{host})-E_{inter}$ (in eV).
Negative values indicate higher stability of the substitutional
configuration over the interstitial. The relative energy of
substitutional and interstitial Mn in pure Si or Ge are included for
comparison.\label{tab:energetics}}
\begin{ruledtabular}
\begin{tabular}{lccc}
&P&As&Undoped\\
\hline
Si&$-0.84$&$-0.87$&$+0.58$\\
Ge&$-1.35$&$-1.42$&$-0.63$\\
\end{tabular}
\end{ruledtabular}
\end{table}

\begin{figure}[b]
 \includegraphics[width=3.2in]{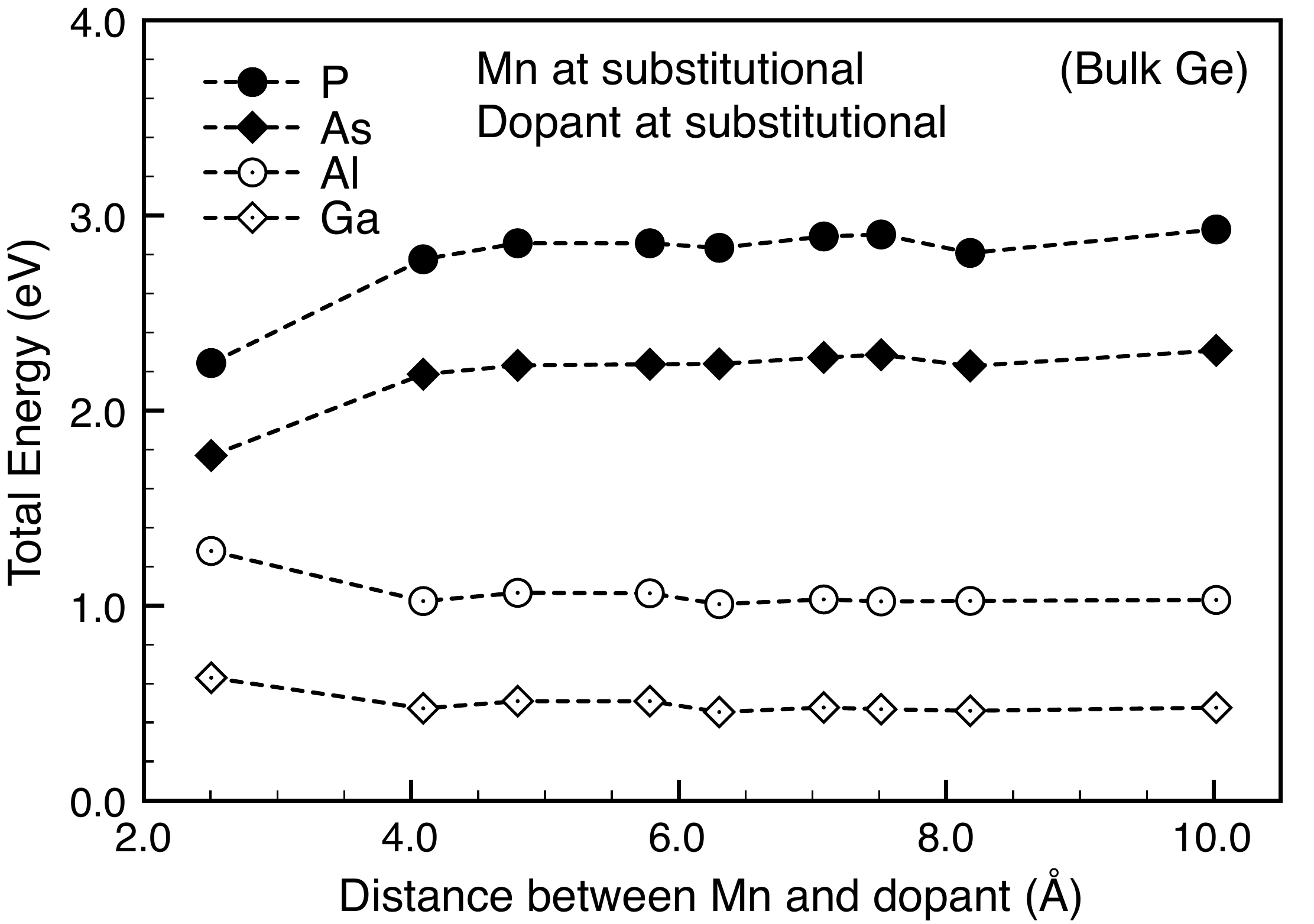}
 \caption{Relative total energy as a function of the distance between a substitutional Mn and a substitutional e-dopant in bulk Ge.\label{fig:s-s-distance}}
\end{figure}

We next calculate the total energy of a Ge-supercell doped by a substitutional Mn/e-dopant pair
at different  separations. The trend of the total energy with increasing distance
between the Mn atom and the e-dopant is shown in Fig.~\ref{fig:s-s-distance}.
The interaction between Mn and e-dopant is attractive for
\textit{n}-type e-dopants (P and As) and repulsive for \textit{p}-type e-dopants (Al and Ga).
This suggests that the picture of electrostatic interaction between Mn and e-dopants that we proposed at the beginning of this section is valid.

Finally we note that the energy differences between the initial and final
states depend on the Mn concentration, as illustrated in Table I. The
calculated $\Delta$E values at the 0.46\% Mn
concentration are larger than those at the 1.56\% concentration, but
for the important cases of \textit{n}-type dopants in Ge, these energy
differences are still much lower than the threshold of
$\sim 0.8$ eV. This relatively strong dependence is not due
to constant volume calculations, because it is also observed when the
supercell volume is fully relaxed. Instead, it is caused by the
interaction between the Mn atoms in adjacent supercells. We stress that
the qualitative picture that the \textit{n}-type dopants facilitate
substitutional incorporation of Mn is valid for all the experimentally
accessible Mn concentrations considered here.

In short, we have shown that in the presence of a neighboring \textit{n}-type dopant,
the substitutional sites are energetically preferred by Mn atoms to interstitial sites
and are kinetically accessible.
In the following sections we will turn to study the electronic and magnetic properties
of this new {\textit{n}-\textit{p} co-doped system.

\section{\label{sec:magnetism}Electronic and magnetic properties of the co-doped systems}

\subsection{Electronic structure}

\begin{figure}[b]
 \begin{center}
\includegraphics[width=3.2 in]{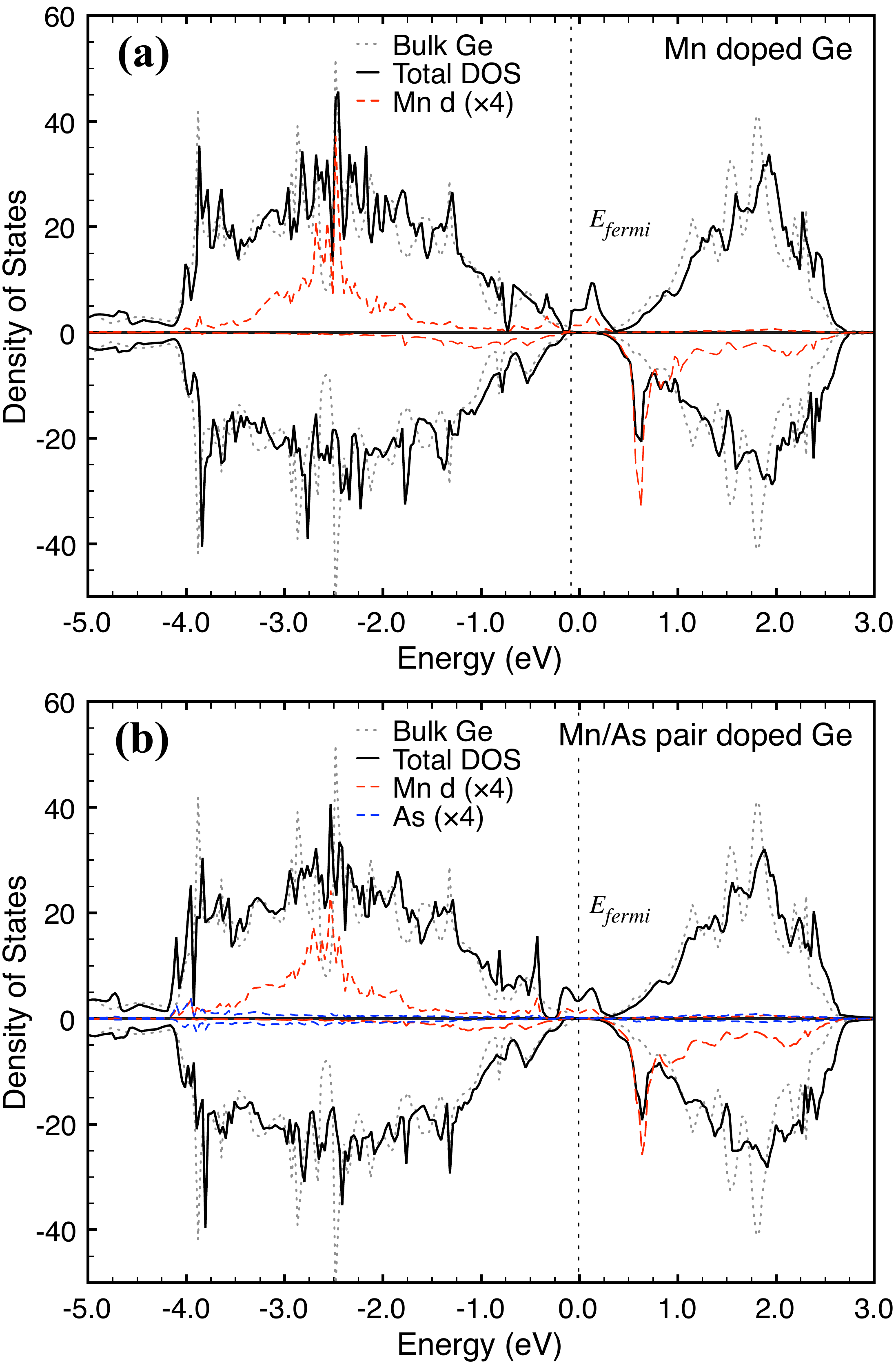}
 \end{center}
 \caption{\label{fig:dos1}\textbf{(color online)} The spin-resolved DOS of (a) Mn doped Ge, and (b) a Mn/As pair doped Ge. Projected DOS of Mn 3$d$ states and As are given as red and blue dashed lines, respectively. The DOS for bulk Ge is shown for comparison.}
\end{figure}

The electronic properties of the Mn/e-dopant co-doped systems
are conveniently presented through the calculated density of states (DOS).
Fig.~\ref{fig:dos1} and Fig.~\ref{fig:dos2} show the total DOS and local DOS for
the substitutional Mn and e-dopants in Ge and Si, respectively.
Several important features emerge: \\
(1) Mn doped Ge or Si are all half metals, regardless of the existence
of e-dopants like As or P, which means the value of the total magnetic
moment per Mn atom is integer.\\
 (2) From the figures it can be determined that the moment per Mn is 3$\mu_{B}$
 in pure Ge or Si, and 4$\mu_{B}$ after co-doping with another e-dopant.
 The importance of this finding, namely, co-doping can actually increase the
 magnetic moment of Mn, will be discussed in the next subsection.\\
(3) The local DOS for Mn is broadened to the whole range of the host valence band,
indicating that there is strong hybridization between the Mn
$d$-state and the valence $p$-state of the host semiconductor.\\
 (4) The local DOS of the additional e-dopant is negligible,
 meaning that the states it contributes are mostly delocalized,
 so that its most evident influence on the total DOS is simply to shift
 the Fermi energy to a higher value, as is expected for a regular non-magnetic dopant.

\begin{figure}[b]
 \begin{center}
\includegraphics[width=3.2 in]{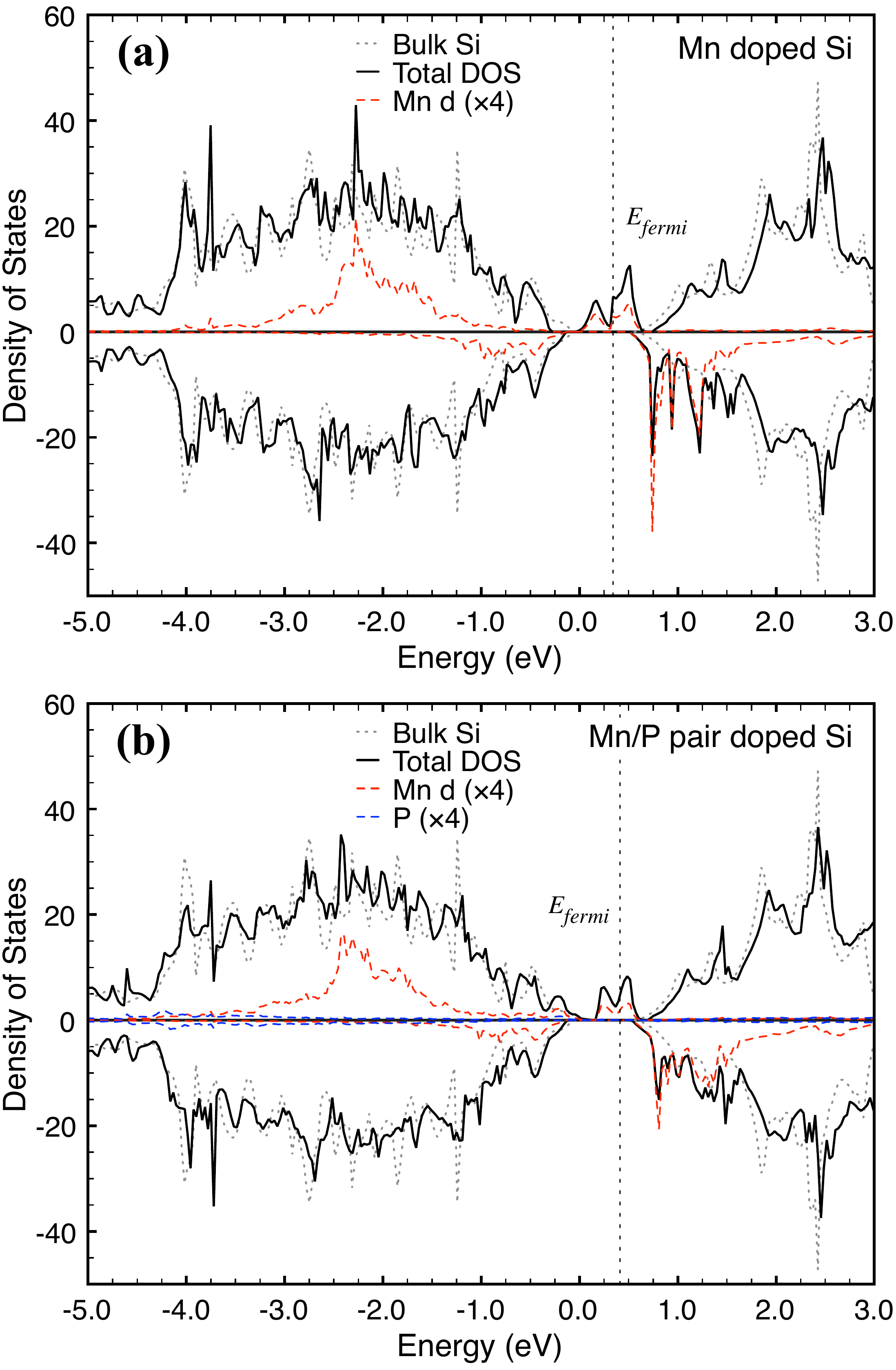}
 \end{center}
 \caption{\label{fig:dos2}\textbf{(color online)} The spin-resolved DOS of (a) Mn doped Si, and (b) a Mn/P pair doped Si. Projected DOS of Mn 3$d$ states and P are given as red and blue dashed lines, respectively. The DOS for bulk Si is shown for comparison.}
\end{figure}

\subsection{Magnetic properties}

At first sight, the additional \textit{n}-type dopant may negatively
influence the strength of the magnetic interaction between Mn atoms,
because of the compensation of hole carriers \cite{jungwirthrmp06,
ohnoscience98, dietlscience00}.  This argument may not be
true for the following reason: We can write the magnetic interaction energy between
two Mn ions as:
\begin{eqnarray}
E_{Mn_1  - Mn_2 }  = J_{eff} S_{Mn_1 }  \cdot S_{Mn_2 },
\end{eqnarray}
where $J_{eff} $ denotes the effective magnetic
coupling strength and $S_{Mn_1 } $,
$S_{Mn_2 } $ represent the local magnetic moments
associated with the Mn atoms; even if $J_{eff} $ were to decrease
because of the carrier compensation effect, since the influence of the
e-dopant on the local moments of Mn is positive
as mentioned in point (2) of the previous subsection,
it is still possible that the enhancement of Mn moments by the e-dopants
outweighs its negative influence on $J_{eff}$.

To check whether this is the case, we first resort to direct \textit{ab initio}
calculation of the magnetic coupling energy of a Mn-Mn pair with different separations, which
can be represented by the total energy difference $\Delta E$ between
the antiferromagnetic (AFM) and the ferromagnetic (FM) states of the pair
\cite{zhaoprl03, stroppaprb03, wengprb05, mahadevanprl04}.
In the present case, each Mn atom has an \textit{n}-type dopant neighbor,
which leads to more spatial configurations with the same Mn-Mn
distances. To be precise, a substitutional Mn atom has four nearest
neighbors, that is, four possible sites for the \textit{n}-type dopant atom, and
thus there are 16 possible configurations for a given
Mn-Mn distance.  The number
of nonequivalent configurations for each of 12 Mn-Mn separations in
the range 2.4 - 9.5 \AA~ in Si (2.5 - 10.0 \AA~ in Ge) is 2, 6,
16, 4, 7, 16, 4, 7, 16, 10, 7 and 5 respectively, with increasing distance.

\begin{figure}
 \begin{center}
\includegraphics[width=3.2 in]{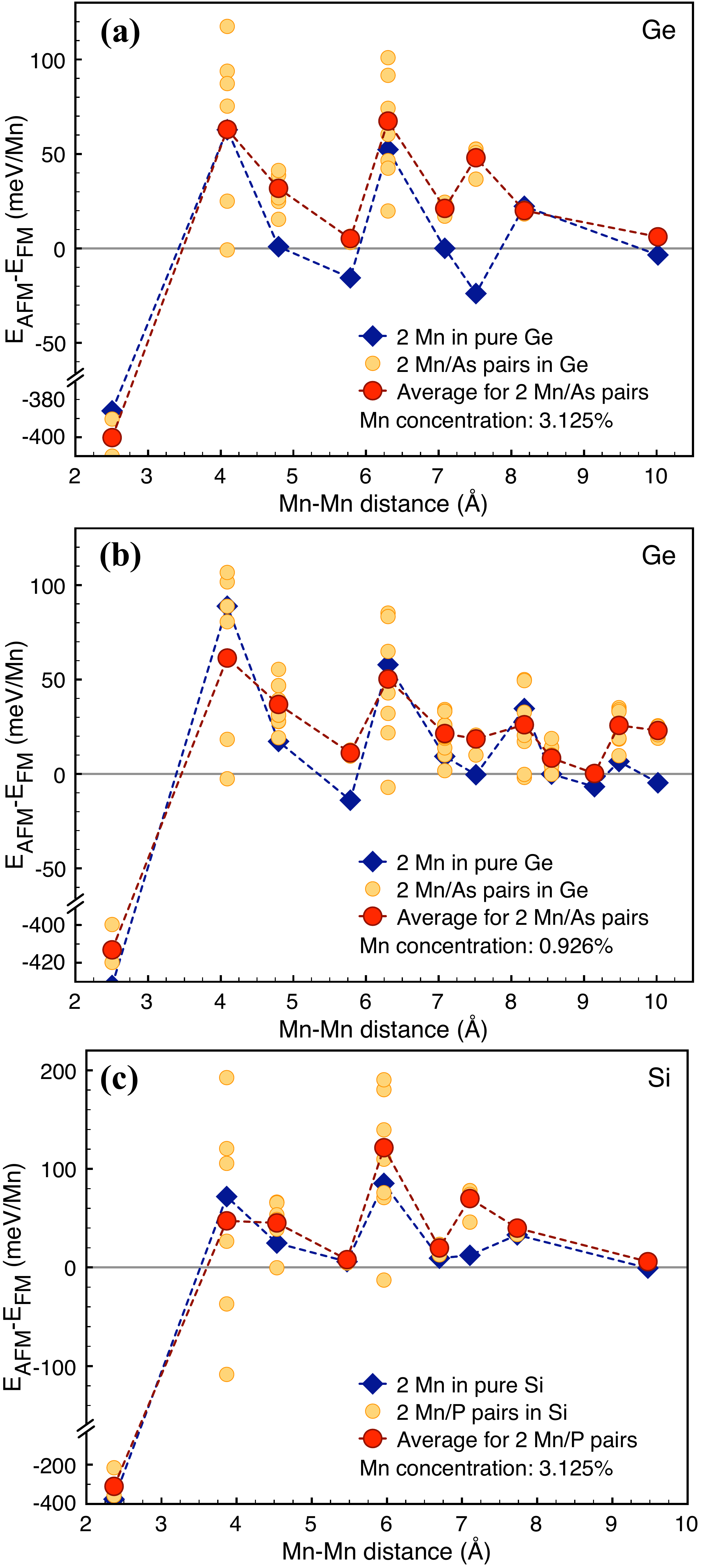}
 \end{center}
 \caption{\textbf{(color online)} Total energy difference between AFM and FM states of two Mn
 ions versus Mn-Mn separation for two Mn/As pairs doped Ge at (a) 3.125\% and
 (b) 0.939\% Mn concentration;
 (c) for two Mn/P pairs doped Si, represented by small orange dots.
 The large red dots are averages over the small orange dots for a given Mn-Mn distance. For comparison, the results for the systems doped with only two Mn impurities are shown as blue diamonds.\label{fig:coupling}}
\end{figure}

\begin{figure}
 \begin{center}
\includegraphics[width=3.2 in]{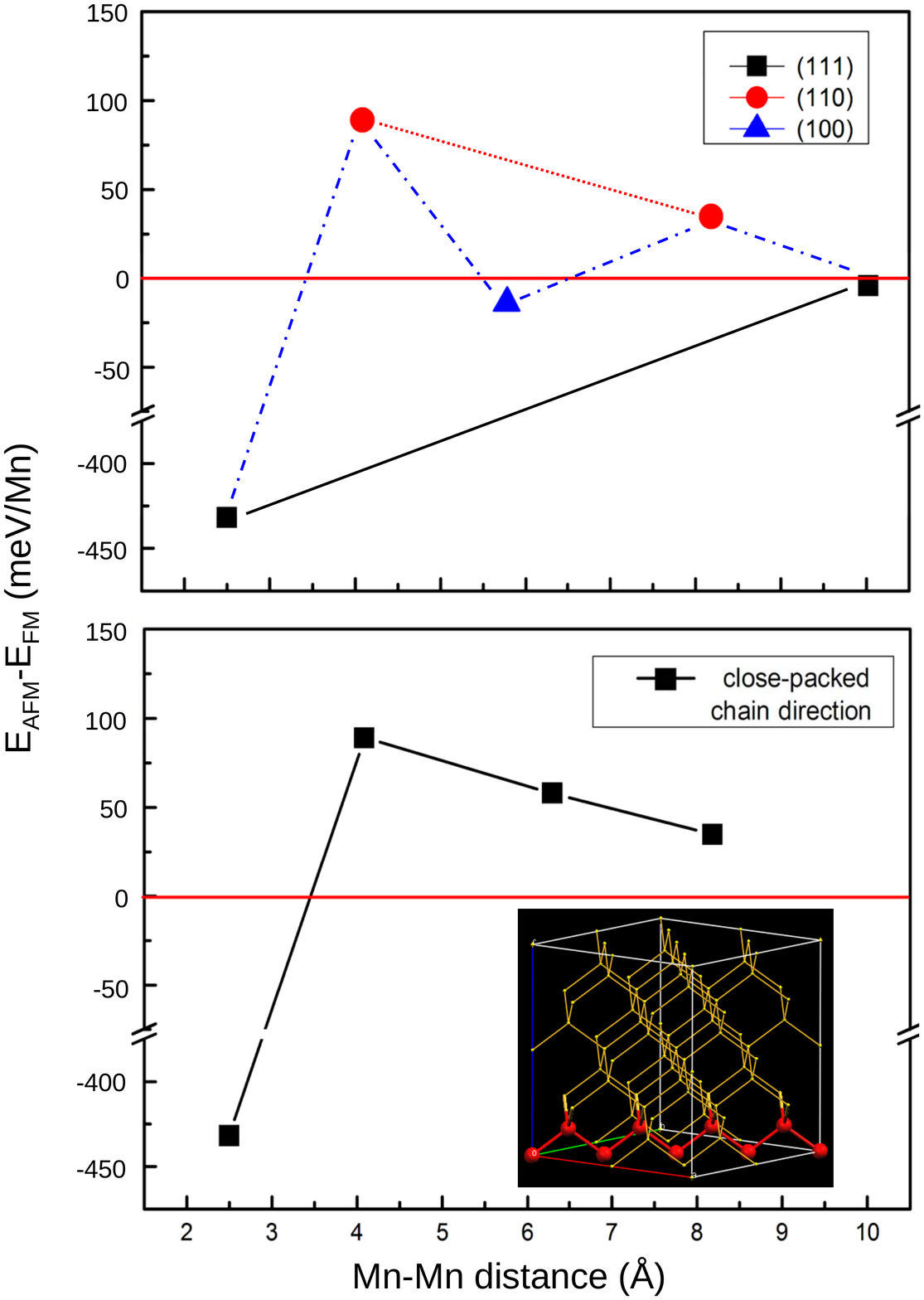}
 \end{center}
\caption{\textbf{(color online)} Magnetic coupling along different crystal directions.
Black squares are for the (111) direction, red dots for the (110) direction and blue triangles
for the (100) direction. Magnetic coupling along a close-packed chain is illustrated in the inset.\label{fig:coupling-direction}}
\end{figure}

With these considerations, our results for Mn/As co-doped Ge and
Mn/P co-doped Si are shown in Fig.~\ref{fig:coupling}.
We first note that in the case of two Mn atoms in pure Ge ((a) and (b) of Fig.~\ref{fig:coupling}),
the behavior of the AFM-FM energy difference $\Delta E$ is oscillatory
between positive and negative values as a function of distance
(but monotonically decreasing along different directions, see below). In contrast, the
average interaction between the two Mn/As pairs in Ge always
favors FM coupling except at the nearest-neighbor Mn-Mn distance, and this characteristics
does not change with doping level (compare Fig.~\ref{fig:coupling}(a) of 3.125\% Mn
and Fig.~\ref{fig:coupling}(b) of 0.926\% Mn). In the case of Si as host, Mn atoms favor FM coupling except for nearest neighbor distance, and this feature does not change upon co-doping.

Though usually it is assumed that the hole-mediated magnetic interaction in dilute magnetic semiconductors is RKKY-like, we find that the FM-AFM oscillation displayed
in Fig.~\ref{fig:coupling} should not be treated as a manifestation of the RKKY interaction~\cite{zhaoprl03, stroppaprb03, wengprb05, mahadevanprl04}.
Fig.~\ref{fig:coupling-direction}(a) shows the magnetic coupling between
Mn ions along different lattice directions, which is similar to the design in Mahadevan's work~\cite{mahadevanprl04}, where it is evident that the oscillatory behavior
is replaced by monotonic decrease in magnitude.
On the other hand, for the doping levels considered here,
the period of RKKY oscillation is much larger than the lattice constant~\cite{matsukuraprb98,kudrnovskyprb04}, as in the case of GaAs.
Thus, the oscillation here is merely due to magnetic anisotropy,
rather than a manifestation of RKKY-type interaction.
In Fig.~\ref{fig:coupling-direction}(b), we plot the coupling
along a close-packed atom chain, which also shows monotonic decrease with distance
except for the nearest neighbor value.
This finding leads us to suggest that Mn ions in Ge are magnetically
coupled through some paths consisting of covalently-bonded Ge atoms,
a hypothesis which deserves to be checked by more detailed investigations.

There are two other important issues revealed in Fig.~\ref{fig:coupling}.
First, the dispersive values of $\Delta E$ at a
given distance in the co-doped case show a new kind of
magnetic anisotropy. This anisotropy is
conceptually different from the magnetic anisotropy that is
typically discussed in the literature. The traditional definition
refers to the cases in which the magnetization of a system exhibits
anisotropy when the magnetic moment is polarized along different
crystalline directions \cite{wangprl93}, or when the coupling
between two magnetic dopants is anisotropic depending on their
relative orientation in the host material \cite{mahadevanprl04}. In
contrast, here the two magnetic dopants are fixed in space,
and the magnetic anisotropy is caused by the relative positions of
the two \textit{n}-type e-dopants surrounding the
magnetic impurities. Second, from the presence of the \textit{n}-type dopant,
the FM interaction between two magnetic atoms on the whole preserves its
magnitude rather than being substantially weakened. Thus, the influence
of e-dopants on the magnetic properties of the whole system is not simply
a weakening of the magnetic coupling by decreasing the number of interaction mediators.

Deeper understanding of the above observations requires a careful
examination of the microscopic coupling mechanism. To this end, we
consider three representative configurations, all with the same Mn-Mn
separation fixed, equal to the next nearest neighbor distance in the
Ge matrix: (a), a Mn-Mn pair in pure Ge as the reference structure; (b) and (c), a
Mn/e-dopant-Mn/e-dopant pair with the strongest and weakest magnetic
couplings, respectively.
For the reference case shown in Fig.~\ref{fig:spindensity}(a), the two Mn atoms share a Ge
atom as their nearest neighbor. When the Mn pair is ferromagnetically
coupled, the spin density in the plane containing the two Mn atoms and
their mutual Ge neighbor (the (1 $\overline{1}$ 0) plane indicated in Fig.~\ref{fig:spindensity}(a)) is
plotted in Fig.~\ref{fig:spindensity}(b). The red (blue) area represents spin up (down)
density. The large local magnetic moments of Mn induce
spin polarization on the nearby non-magnetic Ge atoms, which are
antiferromagnetically coupled with the Mn atoms.

\begin{figure}
 \begin{center}
\includegraphics[width=3.4 in]{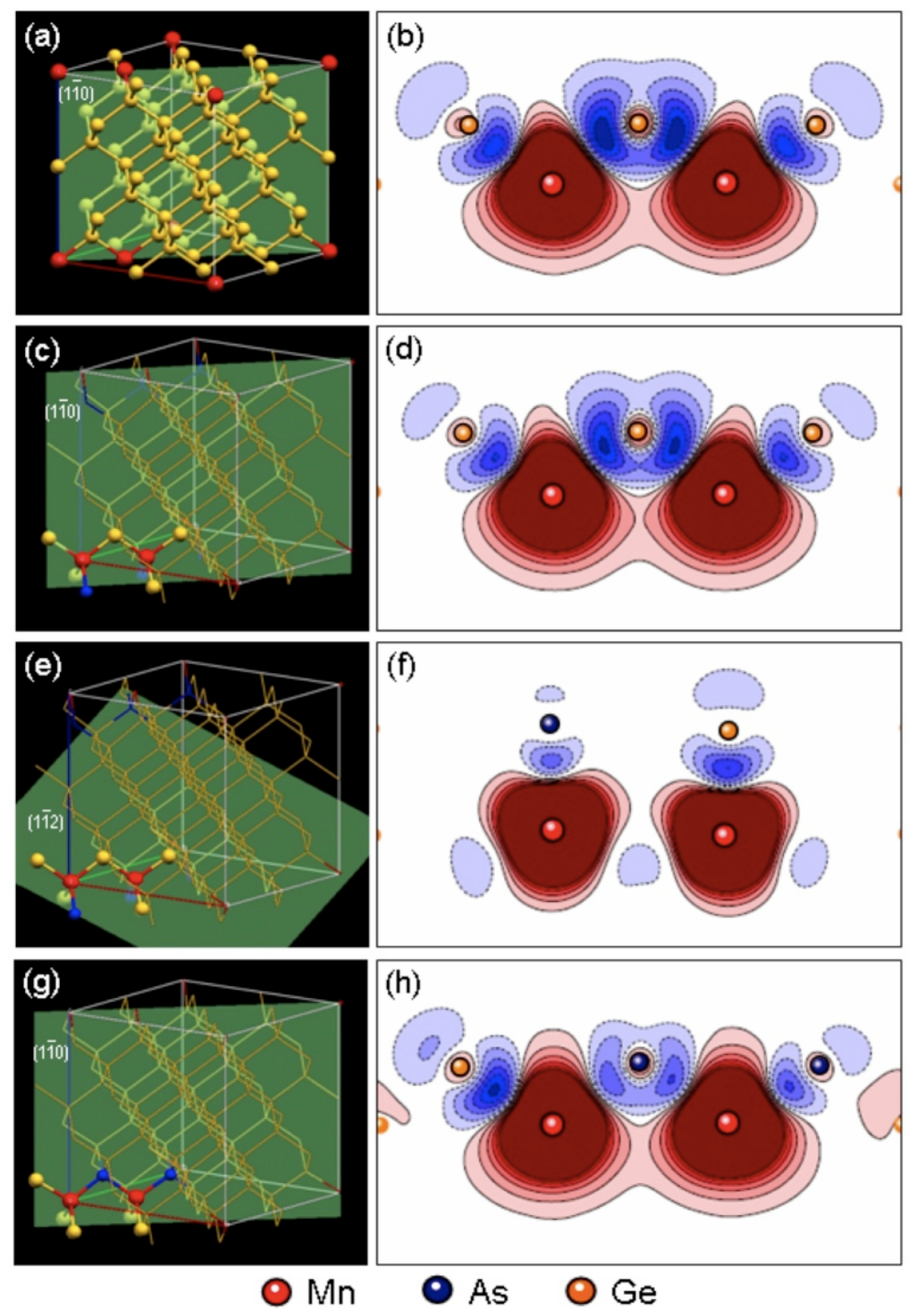}
 \end{center}
 \caption{\textbf{(color online)} The atomic structures and spin density plots of
 three representative configurations of two Mn TOMS, (a) and (b),
  and two Mn/As pairs, (c)-(h), doped Ge.
  In all structures, the two Mn atoms are fixed at the next nearest neighbor distance.
  The spin density plots are taken on the green plane as depicted in the structures on the left.
  The red and blue contours represent the two different spin components.
  (c) and (e) correspond to the configuration with the strongest magnetic coupling
  between two Mn ions, and (g) the weakest magnetic coupling.\label{fig:spindensity}}
\end{figure}

The corresponding plots for case (b) are shown in Fig.~\ref{fig:spindensity}(c)
and Fig.~\ref{fig:spindensity}(d). In
this case, the two As atoms are not in the (1 $\overline{1}$ 0) plane, and the two
Mn atoms still have the same Ge atom as their mutual nearest neighbor.
Furthermore, the local magnetic moment of the bridging Ge atom shows
little change, indicating that \textit{J$_{eff}$} essentially stays the
same. To show the effect of As doping, we plot in Fig.~\ref{fig:spindensity}(f) the spin
density on the plane containing the two Mn and one As atom (the (1 $\overline{1}$
2) plane in Fig.~\ref{fig:spindensity}(e)). Here, As acts as a donor helping to
compensate the holes
introduced by its neighboring Mn, resulting in an increased local magnetic
moment on each Mn atom, \textbf{$S$$_{Mn}$}
(3.60$\mu{}$$_{B}$$\rightarrow{}$4.00$\mu{}$$_{B}$).
Therefore, the overall magnetic coupling between the two Mn atoms is
enhanced relative to the pure Ge case.

For case (c), the corresponding plots are shown in Fig.~\ref{fig:spindensity}(g) and Fig.~\ref{fig:spindensity}(h).
In this case, the two As and two Mn atoms are both in the (1 $\overline{1}$ 0)
plane, with one As replacing the mutual nearest Ge neighbor of the two
Mn atoms. Similar to case (b), here both
\textbf{$S$$_{Mn1 }$}and\textbf{$S$$_{Mn2}$} are also increased
(\textbf{$S$$_{Mn1}$}:3.60$\mu{}$$_{B}$$\rightarrow{}$3.87$\mu{}$$_{B}$;
\textbf{$S$$_{Mn2}$}:3.60$\mu{}$$_{B}$$\rightarrow{}$4.02$\mu{}$$_{B}$;
the asymmetry in the increase is caused by the asymmetric locations of
the two As atoms). However, because the local magnetic moment of the
bridging atom is substantially decreased from that of case (a) (Ge:$-0.16 \mu{}$$_{B}$$\rightarrow{}$As:$-0.05 \mu{}$$_{B}$), the corresponding
\textit{J$_{eff}$} is also significantly weakened, leading to an
overall weakened magnetic coupling between the two Mn atoms relative to
the pure Ge case.

Summarizing, As as \textit{n}-type dopant can enhance the local magnetic moments of
neighboring Mn atoms, but itself is weakly spin polarized (much weaker
than Ge). Therefore, if As serves as the bridging atom between two Mn
atoms, the global magnetic coupling will be weakened. If As is located
so as to only enhance the magnetic moment of Mn, with a Ge atom still
bridging the Mn-Mn coupling, then the global magnetic coupling will be
enhanced. This conclusion is further confirmed by checking other Mn-Mn
distances.

\section{\label{sec:tc} Curie temperature}

To study the macroscopic magnetic properties of the co-doped DMS
materials using our \textit{ab initio} results, we turn to the
classical Heisenberg model:
\begin{eqnarray}
H =  - \sum\limits_{i,j} {J_{ij} \sigma _i  \cdot \sigma
_j },
\end{eqnarray}
where $J_{ij} $ is the magnetic coupling constant
between moment $i$ and $j$, and
$\sigma _i $ is a unit vector representing the
direction of spin $i$. Then the AFM-FM energy difference
$\Delta E$ calculated in previous section is given by:
\begin{eqnarray}
\Delta E = E_{AFM}  - E_{FM}  = 4J_{12},
\end{eqnarray}
with 1 and 2 the indices of the two moments in the supercell.
With given coupling parameters, we then use Monte Carlo simulations to address the statistical mechanics of the
DMS systems at finite temperatures. To eliminate finite size
effects, the cumulant crossing method \cite{binder} is used to
determine the Curie temperature. This two-step approach has the
distinct advantage over the ordinary mean field approach, that both
disorder and percolation effects are naturally and precisely taken
into account \cite{satoprb04,bergqvistprl04}.

A subtle issue in the present case is the following: Since real interactions
between magnetic atoms in DMS have a built-in
multiatom nature, in an optimal Heisenberg description the coupling
parameters must depend on the system geometry. This is
difficult, if not entirely impossible\cite{franceschettiprl06,franceschettijpcm07} to
address, because of the very large number of possible configurations
in a macroscopic system and some approximations are necessary.

The supercell
\textit{ab initio} approach \cite{sanvitoprl01,sandratskiiprb03,dasilvajpcm04,wierzbowskaprb04,picozzinjp08}
employed here assumes only one approximation,
the pair-superposition approximation, which means that the interaction is
exclusively pairwise and can be added independently to get the total
interaction. Though this may not hold at high
concentrations of magnetic moments, we claim that it should be a
reasonable approximation at the low concentrations we considered
(3.13\%-6\%), where the average distance between two Mn atoms,
calculated by
\begin{eqnarray}
\bar d = 2\sqrt[3]{{\frac{3}{{32\pi x}}}}a
\end{eqnarray}
with $x$ the concentration and
$a$ the lattice constant, ranges from
1.97$a$ to 1.58$a$, or
11.39\AA{} to 9.14\AA{} in the case of Mn doped Ge. Considering the bond length $d_{bond}$ in Ge is only about 2.5\AA{}, a separation $\sim 4d_{bond}$ is large enough for the system to be treated in this approximation.

Using the \textit{ab initio} coupling parameters for Mn$_x$Ge$_{1-x}$,
we first find that MC does not yield identifiable T$_{c}$ up to $x=6\%$ (see below). Nevertheless, after co-doping with As, MC shows that the system has high T$_{c}$, as summarized in Fig.~\ref{fig:tc}, in which we also
include the results from the mean field approximation (MFA) using the
formula \cite{bergqvistprb05}
\begin{eqnarray}
T_c  = \frac{1}{{k_B }} \cdot \frac{{2x}}{3}\sum\limits_{i
\ne 0} {J_{0i} } . \label{mfatc}
\end{eqnarray}

\begin{figure}
 \begin{center}
\includegraphics[width=3.2 in]{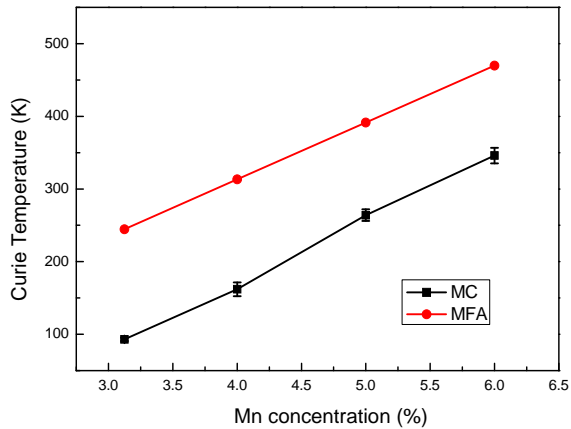}
 \end{center}
 \caption{\textbf{(color online)} Comparison between Curie temperatures calculated by
 the Monte Carlo approach (MC, black squares)
 and those obtained by the mean field approximation (MFA, red dots).\label{fig:tc}}
\end{figure}
These results show that the MFA greatly overestimates the Curie
temperature, as established before~\cite{bergqvistprl04,satoprb04,bergqvistprb05}.
At $x=5\%$, T$_c$ is evaluated to be 264K through MC,
which is much higher than the 118K of 5\% Mn doped GaAs \cite{edmondsapl02}.
At the 6\% Mn concentration, MC gives a T$_{c}$
higher than room temperature. Considering that $x$=6\% is already a
relatively high concentration, we expect that the
pair-superposition approximation may not be valid in this case.
Arsenic doping can still be expected to dramatically change the magnetic properties
of Mn doped Ge, namely, from no finite T$_c$ to a potentially high T$_c$ DMS material.

The dependence of the Curie temperature on Mn concentration, as obtained from
the MC results, is almost linear.
This behavior is partly due to the pair-superposition approximation we used,
meaning that the strength of magnetic coupling does not depend on Mn concentration.
The only influence of concentration on T$_c$ is the average number of magnetic impurity
atoms on each coordination shell. Thus, after the configurational average,
we expect that the dependence of T$_c$ on $x$ resembles the linear one obtained
within the MFA (Eqn.~\ref{mfatc}).
Another reason for this linearity is that the concentrations we studied
are higher than the magnetic percolation threshold of this system \cite{bergqvistprb05}.

\begin{figure}
 \begin{center}
\includegraphics[width=3.2 in]{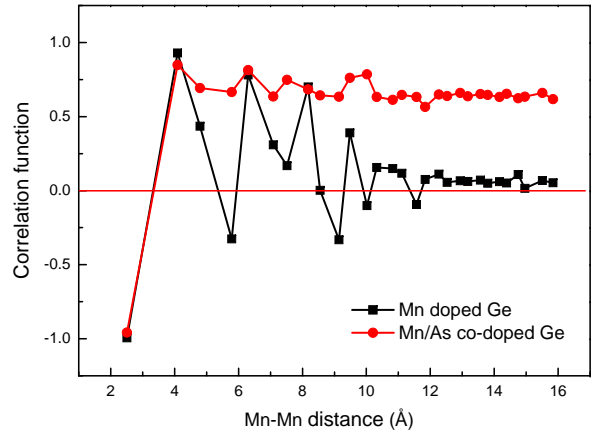}
 \end{center}
 \caption{\textbf{(color online)} Spin-spin correlation function obtained from the
  Monte Carlo simulations. \label{fig:correlation}}
\end{figure}

The presence of AFM couplings and the absence of T$_c$ in the case of pure Mn
doped Ge suggests the possibility of a spin-glass ground state for this system.
Jaeger {\it et al.}~\cite{jaegerprb06} claimed that at low temperatures
Mn$_{x}$Ge$_{1-x}$ exhibits spin-glass-like behavior and the critical temperature
of the spin-glass phase transition is 12K and 15K, for Mn concentrations $x=0.04$ and $x=0.2$,
respectively. To examine whether this is the case,
we first study the spin-spin correlation function of 5\% Mn doped Ge, at $T=0.01K$.
The result is shown in Fig.~\ref{fig:correlation}, along with
a plot for Mn/As co-doped Ge, for comparison.
The correlation function of Mn$_{x}$Ge$_{1-x}$ decays very fast with increasing distance
and approaches to zero, indicating the absence of FM order
even at low temperatures. Using the spin-glass order parameter, defined as
\begin{eqnarray}
q=\frac{1}{N}\sum\limits_i \langle \mathbf{S}_i \rangle ^2, \label{sgorder}
\end{eqnarray}
and a similar cumulant crossing method~\cite{binder} yields a transition
temperature $\sim$5K, a value in semi-quantitative agreement with the results
of Jaeger {\it et al.}~\cite{jaegerprb06}.

\section{\label{sec:discussion} Discussion}

Mn$_{x}$Ge$_{1-x}$ has been attractive within the DMS community because of its easy incorporation into the current semiconductor industry.
The mechanism of valence hole mediated ferromagnetism for (Ga,Mn)As was proposed years
ago~\cite{matsukuraprb98,ohnojmmm99,dietlscience00} and has been extensively accepted ever since, but there is still no definitive
theory for Mn$_{x}$Ge$_{1-x}$\cite{jungwirthrmp06}.
One reason, which is also one of the main points of this paper,
is the difficulty of decreasing the percentage of interstitial Mn dopants.
The other important point is the hard-to-control inhomogeneity of this system,
which has been realized only in recent years.
The high Curie temperature formerly reported in
Mn$_{x}$Ge$_{1-x}$\cite{parkscience02,choprb02}
is now thought to be due to the formation of Mn-rich
regions in the host semiconductor\cite{jungwirthrmp06,liapl05,liprb05,jaegerprb06,dietljap08}.
For example, Mn-rich nanodots\cite{bougeardprl06} and nanocolumns\cite{kangprl05,jametnm06,liprb07} in Mn$_{x}$Ge$_{1-x}$ have been
reported by many experimental groups, and later reproduced in Monte Carlo simulations\cite{katayamapssa07}.

Despite the seemingly unavoidable precipitation or spinodal decomposition\cite{dietljap08}
present during the growth of Mn$_{x}$Ge$_{1-x}$ samples,
the study on homogeneously doped Mn$_{x}$Ge$_{1-x}$ has never stopped.
Work by Li \textit{et al.}\cite{liapl05,liprb05} indicates that the long-range
FM order in Mn$_{x}$Ge$_{1-x}$ only exists at low temperatures ($\leq 12$K).
Jaeger \textit{et al.}\cite{jaegerprb06} claimed that even at low temperatures
Mn$_{x}$Ge$_{1-x}$ shows spin-glass-like behavior, and proposed that this is
 due to the intercluster frustration between FM Mn-rich clusters.
 Recently, Zeng \textit{et al.}\cite{zengprl08}, using a newly developed
 subsurfactant epitaxy method, successfully grew cluster-free Mn$_{x}$Ge$_{1-x}$
 samples with a Mn doping level of 0.25\%. Surprisingly, this low doping level (by normal DMS standards, where 1\% to 5\% is typical) led to a Curie temperature as high as over 400 K.

The results in the present work provide a viewpoint that may resolve
the seemingly conflicting experimental results discussed above.
Specifically, we showed that
the magnetic coupling between Mn ions in Mn$_{x}$Ge$_{1-x}$
oscillates between FM and AFM with increasing Mn-Mn distance
and that homogeneous Mn$_{x}$Ge$_{1-x}$ exhibits spin glass behavior. Thus,
 the FM order observed in experiments could be due to spatially ordered structures,
 which are formed due to precipitation or spinodal decomposition.
 The high transition temperatures are expected because of the large
 concentration of magnetic moments within the clusters. On the other hand,
 the AFM frustration in this case only manifest itself in the inter-cluster interaction,
 and thus leads to the spin-glass behavior at low temperatures.

The unexpected high T$_c$ in Zeng's work requires more discussion.
Upon co-doping with \textit{n}-type dopant As, the AFM coupling
between Mn ions is absent, and a high Curie temperature emerges.
We thus speculate that the high transition temperature in this case originates
from this co-doping effect, and the unexpected \textit{n}-type e-dopant here is most probably oxygen.
Indeed, a recent study on the role of oxygen defects in Mn$_{x}$Ge$_{1-x}$ by Continenza and Profeta\cite{continenzaprb08} supports this scenario, namely that oxygen acts
as an \textit{n}-type dopant and facilitates the substitutional Mn doping.
It is also reasonable to expect a positive influence of oxygen on the Mn-Mn magnetic coupling,
which, together with the possible existence of Mn-rich regions,
can lead to a high Curie temperature.

Recently, the works of Kuroda\cite{kurodanm07} and Bonanni\cite{bonanniprl08} demonstrated
experimentally that the aggregation of magnetic ions in DMS systems can be controlled
by modifying the charge states of the magnetic dopants.
This is in agreement with the spirit of our work, that is, charge states of impurity dopants
play an important role in the growth kinetics of DMS materials
and can lead to different structures with their own specific properties.

Finally, this work suggests that the enhancement of substitutional Mn concentration in group-IV DMS can be achieved in epitaxial growth by co-depositing with the e-dopants. More specifically, this co-doping method can be integrated in the recently developed subsurfactant epitaxial growth\cite{zengprl08}, where pure Ge layers epitaxially grow on a Ge(100) substrate pre-covered with a submonolayer of Mn. During the growth process, the Mn atoms tend to diffuse upward to the subsurface layer, as predicted in a previous theoretical study\cite{zhuprl04}. When the growth is slow enough, a small fraction of the Mn atoms can be trapped in substitutional sites, which leads to homogeneous substitutional Mn doping with all the interstitial Mn floating at the subsurface layer. However, the resulting Mn concentration is still pretty low (0.25\%). Here we propose that by co-depositing Ge with another \textit{n}-type e-dopant, with very low depositing rates, the growth front could mimic the subsurfactant growth mode, but with more efficient substitutional trapping of Mn. The Mn trapping rate can be controlled by changing the concentration of the e-dopant. Experimental confirmation of this co-doping scheme is highly desirable.

\section{\label{sec:conclusion} Conclusion}

In conclusion, our \textit{ab initio} DFT calculations show that in DMS materials additional
\textit{n}-type electronic dopants can serve to enhance the substitutional doping of
\textit{p}-type magnetic dopants such as Mn in the host group IV
semiconductors Si and Ge.
The additional dopants suppress to a large extent
the charge and magnetic-moment compensating effects from interstitial Mn,
which is detrimental to FM order.
We calculate the magnetic coupling between moments associated with Mn atoms
using the energy difference between parallel and antiparallel aligned pairs of Mn moments.
We examined the unconventional magnetic anisotropy in Mn/As co-doped Ge,
namely, the dependence of magnetic coupling on the relative positions
of magnetic ions and their neighboring assistant dopants. We find that the
coupling oscillates between ferromagnetic (FM) and antiferromagnetic (AFM)
with increasnig Mn-Mn distance in the Mn-doped Ge,
whereas in As/Mn \textit{n-p} co-doped Ge the
coupling values at Mn-Mn separations up to the 12th coordination shell
are all FM, except for the nearest-neighbor one.
Moreover, we find that the FM-AFM oscillatory behavior in Mn$_{x}$Ge$_{1-x}$ is due
to anisotropy rather than being the result of a RKKY-type interaction.
Our Monte Carlo simulations, using magnetic coupling parameters
obtained from the \textit{ab initio} calculations, indicate a high
Curie temperature in Mn/As-Ge of 264K at 5\% Mn doping.
On the other hand, no FM order is observed in Mn$_{x}$Ge$_{1-x}$
(without co-doping) as Mn concentration ranges from 3.13\% to 6\%.
Thus, the homogeneously doped Mn$_{x}$Ge$_{1-x}$ is most likely
a generic spin glass, with a spin-glass transition temperature
of ~5K at 5\% doping, also obtained from our Monte Carlo simulations.
Accordingly, we suggest that the high Curie temperature observed experimentally
in Mn$_{x}$Ge$_{1-x}$ is either due to the formation of Mn-rich
spatially ordered regions, or to \textit{n-p} co-doping effects from the
\textit{n}-type oxygen impurities, or a combination of both.

\begin{acknowledgments}
The authors thank Dr. Rong Yu for helpful discussions and Dr. Kirk H. Bevan for a critical reading of the manuscript. This work was supported in part by NSF grant Nos. DMR-0325218 and DMR-0606485, by DOE grant No. DE-FG02-05ER46209, and in part by the Division of Materials Sciences and Engineering, Office of Basic Energy Sciences, DOE. The calculations were performed at NERSC of DOE and NCCS of ORNL.
\end{acknowledgments}

\bibliographystyle{unsrt}

\end{document}